\documentclass[pra,twocolumn,showpacs,showkeys,preprintnumbers]{revtex4}

\usepackage{graphicx}
\usepackage{dcolumn}
\usepackage{bm}
\usepackage{amsmath}
\usepackage{amssymb}
\usepackage{graphicx}
\usepackage{float}

\newcommand{\ket}[1]{\lvert#1\rangle}

\begin{document}

\title{Multi-photon quantum interference with high visibility using multiport beam splitters}

\author{M. Stobi\'nska}
\affiliation{Institute of Theoretical Physics and Astrophysics, University of Gda\'nsk, ul. Wita Stwosza 57, 80-952 Gda\'nsk, Poland}
\affiliation{Institute of Physics, Polish Academy of Sciences, Al. Lotnik\'ow 32/46, 02-668 Warsaw, Poland}

\author{W. Laskowski}
\affiliation{Institute of Theoretical Physics and Astrophysics, University of Gda\'nsk, ul. Wita Stwosza 57, 80-952 Gda\'nsk, Poland}

\author{M. Wie\'sniak}
\affiliation{Institute of Theoretical Physics and Astrophysics, University of Gda\'nsk, ul. Wita Stwosza 57, 80-952 Gda\'nsk, Poland}

\author{M. \.Zukowski}
\affiliation{Institute of Theoretical Physics and Astrophysics, University of Gda\'nsk, ul. Wita Stwosza 57, 80-952 Gda\'nsk, Poland}

\date{\today}


\begin{abstract}
Multi-photon states can be produced in multiple parametric down conversion (PDC) processes. The nonlinear crystal in such a case is pumped with high power. In theory, the more populated these states are, the deeper is the conflict with local realistic description. However, the interference contrast in multi-photon PDC experiments can be quite low for high pumping. We show how the contrast can be improved. The idea employs currently accessible optical devices, the multiport beam splitters. They are capable of splitting the incoming light in one input mode to $M$ output modes. Our scheme works as a POVM filter. It may provide a feasible CHSH-Bell inequality test, and thus can be useful in e.g.\ schemes reducing communication complexity. 
\end{abstract}

\maketitle

\section{Introduction}

Quantum interference results from the superposition principle, the key feature of the quantum world. The possibility of increasing the average photon population of experimentally accessible quantum states of light and their superpositions, poses an intriguing question: how to observe quantum interference for light on the mesoscopic scale? Multi-photon interference allows to test fundamental aspects of quantum mechanics and finds applications in quantum information processing~\cite{review, review2}. 

Multi-photon states naturally arise in parametric frequency down conversion (PDC) with strong pumping~\cite{pdc}. PDC is the power horse of quantum optics in generation of quantum light. It is a spontaneous process, in which a blue photon from a pumping laser is split by a $\chi^{(2)}$ non-linear crystal into a pair of lower energetic entangled red photons. For the low pump power, to a good approximation, the type-II PDC output is a two-photon polarization singlet state. In this regime, PDC-based light sources provide a very good higher-order quantum interference contrast allowing e.g., to falsify the Leggett-type model with non-local hidden variables~\cite{LeggettNature}. As the power of the pump is increased, more entangled pairs can be created simultaneously from a single pump pulse giving rise to multi-photon states. It is then possible to generate, for example, Dicke states~\cite{dicke1,dicke2}, which were realized up to 6 photons, or bright squeezed entangled vacuum~\cite{singlet,Masha}, containing from few up to $10^{13}$ photons on average. Probing quantum properties of these states is subtle~\cite{Stobinska1,Stobinska2,Stobinska3}. On the one hand, their multi-photon nature puts them in a deeper conflict with local realistic description~\cite{Mermin1990}. On the other, high pumping power fundamentally impairs their interference contrast~\cite{Laskowski2009} (e.g. reduces the Hong-Ou-Mandel dip~\cite{HOM}), making the observation of non-classical phenomena difficult. 

Here we show a method to boost the interferometric contrast. We consider a typical scheme for a Bell inequality test. It involves multiport beam splitters~\cite{ZukowskiHorneZeilinger}. A multiport beam splitter is an optical device, which is capable of splitting the light in one input mode, to $M$ spatial output modes. The simplest example of a multiport is a $50:50$ beam splitter. Multiports can be realized with integrated optics~\cite{Perucco2011}. The idea that lies behind our scheme is related to time-multiplexing in optical fibers~\cite{Fitch,Achilles}, but here we multiplex spatial modes. Effectively, it works as a POVM (positive operator valued measure) filter on the input state and projects it onto a subspace of a two-photon singlet state. 

In Section \ref{Sec1} we introduce the multi-photon states emerging from PDC and analyze their visibilities in terms of two detection schemes: analog detection and single photon counting. In the next two sections we discuss schemes which allow to increase the interference contrast. In Section \ref{Sec2} we discuss an interferometer which employs a beam splitter and a hybrid detection strategy. Section \ref{Sec3} is devoted to an interferometer involving a multiport and the photon counting detection only.

\section{Statistical properties of the PDC radiation versus local realism} \label{Sec1}

Quantum nature of correlations of light may be revealed by quantum interference if its visibility is greater than certain threshold value~\cite{Mandel,Stobinska&KW}. In particular, to show a conflict between classical description and predictions of quantum mechanics for a two-photon singlet state using CHSH-Bell inequality~\cite{CHSH}, visibility (interferometric contrast) in coincidence counts should be greater than the critical value $v_{crit}=1/\sqrt{2}$. It is so far not known what is the exact threshold value for visibility for the multi-photon states to violate a Bell inequality. What is known however, is that all non-classical effects, including violation of any Bell inequality, require high visibility. In multi-photon PDC experiments visibility may be impaired by fundamental properties of the emission process (production of additional pairs of photons) and misalignment~\cite{ZUK-91,Laskowski2009}. The measured value of visibility depends on the applied photodetection technique. In our considerations we take the value of visibility $v_{crit}=1/\sqrt{2}$ as a benchmark of nonclassicality, without claiming the impossibility of local realistic description of the state.

\begin{figure}
\begin{center}
\includegraphics[width=0.28\textwidth]{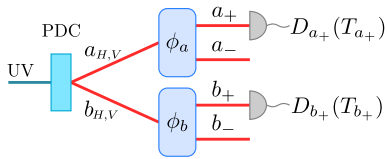}
\caption{Interferometric scheme employed for Bell inequality testing with polarization entanglement and one of the two detection techniques: linear efficiency ($D$) and ``on-off'' detection ($T$).}
\label{setup}
\end{center}
\end{figure}

To illustrate the effect of the deteriorated visibility, let us discuss the interferometer in Fig.~\ref{setup}. This configuration is also a standard Bell-type experiment involving one source. We consider a non collinear PDC process with type II phase matching approximately described by the Hamiltonian
\begin{equation}
\mathcal{H} = i g (a_H^{\dagger} b_V^{\dagger} - a_V^{\dagger} b_H^{\dagger}) + \mathrm{h.c.},
\label{PDC-Hamiltonian}
\end{equation}
where $g$ is the coupling constant proportional to pump power and second-order nonlinearity of the crystal. The photons are emitted in linear polarization H (horizontal) and V (vertical) into two spatial modes described by creation operators $a^{\dagger}$ and $b^{\dagger}$. 
The unitary evolution operator $\mathcal{U}(t) = \exp\{i \mathcal{H} t\}$ gives the following state evolution
\begin{align}
\mathcal{U}(t) |0\rangle &{}= e^{i \mathcal{H} t } |0\rangle = e^{- K (a_H^{\dagger} b_V^{\dagger} - a_V^{\dagger} b_H^{\dagger} - \mathrm{h.c.}) } |0\rangle 
\nonumber\\
&{}= \frac{1}{\cosh^2{K}} e^{\tanh{K} (a_H^{\dagger} b_V^{\dagger})} e^{-\tanh{K} (a_V^{\dagger} b_H^{\dagger})} |0\rangle.
\label{unitary}
\end{align}
The ``disentanglig theorem''~\cite{Sekatski} is used between the first and the second line in Eq.~(\ref{unitary}), and $K=gt$ where $t$ is the effective interaction time in the crystal. If the pump-crystal coupling is weak, i.e. approximately one pump photon out of $10^5$ is successfully down-converted, it is tempting to use the first-order perturbation expansion. The output state in Eq.~(\ref{unitary}) is a superposition of vacuum and two maximally entangled photons
 \begin{equation}
|\Psi\rangle \simeq |0\rangle + K \frac{1}{\sqrt{2}} \left(|1\rangle_{a}^H |1\rangle_{b}^V - |1\rangle_{a}^V |1\rangle_{b}^H\right).
\label{biphoton}
\end{equation}
This approximation leads to statistically independent emissions of pairs of entangled photons. However, for stronger pumping this description turns out to be oversimplified. If the pump power is very high, PDC works as an optimal quantum cloning machine based on optical amplification. The higher-order terms in the expansion of unitary evolution operator are important. The resulting output state is called entangled bright squeezed vacuum~\cite{SIMON-BOUWMEESTER}
\begin{align}
|\Psi\rangle &{}= \frac{1}{\cosh^2{K}} \sum_{n=0}^{\infty} \sqrt{n+1} \tanh^n{K} |\psi^{(n)}_-\rangle,
\label{state}
\\
|\psi^{(n)}_-\rangle &{}= \frac{1}{\sqrt{n+1}} \sum_{m=0}^n (-1)^m|n-m\rangle_{a}^H |m\rangle_{a}^V |m\rangle_{b}^H |n-m\rangle_{b}^V, \nonumber
\end{align}
where $|\psi^{(n)}_-\rangle$ is an analog of a singlet state of two spin-$\frac{n}{2}$ particles. For small gain $K$, Eq.~(\ref{state}) approaches Eq.~(\ref{biphoton}) (only $n,m=0,1$ contribute). It is interesting to note that (see Eq.~(\ref{unitary})) $|\Psi\rangle$ is given by a product of two two-mode entangled squeezed vacua 
\begin{align}
|\Psi\rangle &{}= |\Psi_1\rangle \otimes |\Psi_2\rangle, 
\label{state-product}\\
|\Psi_1\rangle &{}= \frac{1}{\cosh{K}}\sum_{n=0}^{\infty} \tanh^n{K} |n\rangle_{a}^H |n\rangle_{b}^V, 
\nonumber\\
|\Psi_2\rangle &{}= \frac{1}{\cosh{K}}\sum_{m=0}^{\infty} (-1)^m \tanh^m{K} |m\rangle_{a}^V |m\rangle_{b}^H.
\nonumber
\end{align}

In a usual Bell test, the state given in Eq.~(\ref{state}) is subjected to polarization rotations $\phi_a$ and $\phi_b$. In our analysis we shall consider only elliptic polarization measurements for which the annihilation operators $a_{\pm}$ and $b_{\pm}$ of photons in the modes observed by the detectors are expressed in terms of the input annihilation operators as follows
\begin{equation}
a_{\pm}  = \frac{1}{\sqrt{2}}( a_{H}   \pm e^{i \phi_a} a_{V} ), \;
b_{\pm}  = \frac{1}{\sqrt{2}}( b_{H}   \pm e^{i \phi_b} b_{V}).  
\label{measurement}
\end{equation}

In the following subsections, we will calculate values of visibilities for various interference experiments, considering two photodetection techniques.

\subsection{Linear efficiency detection}\label{linear}

\begin{figure}
\centering
\includegraphics[width=0.45\textwidth]{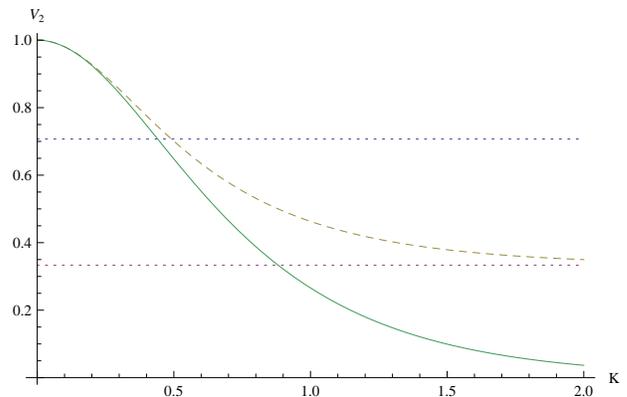}
\caption{Visibilities as functions of amplification gain $K$ for linear efficiency (dashed line) and for ``on-off'' detection (solid line). Horizontal lines represent critical value of visibility $1/\sqrt{2}$, above which violation of CHSH inequality is obtained, and $1/3$, visibility obtained for thermal state.}
\label{fig:v}
\end{figure}

We will start our discussion with linear efficiency devices, i.e. analog detectors. These are PIN-diodes and photomultipliers adjusted to work in such a regime. They are described in terms of the standard photodetection theory, based on the photoelectric effect. Namely, they are measuring electron photocurrent and detector's output is a continuous variable, proportional to the intensity of impinging light. As a result, they give intensity of incoming light integrated over detection time. Probability of joint detection by two detectors $D_{a+}$ and $D_{b_+}$ in Fig.~\ref{setup} equals 
\begin{equation}
 p(D_{a_+}, D_{b_+}| \phi_a,\phi_b) \propto \eta\, \langle \Psi |{:}\, \mathcal{N}_{a_+} \mathcal{N}_{b_+}{:} |\Psi \rangle,
\end{equation}
where $\eta$ is an overall detection efficiency, $\mathcal{N}_{x}$ is photon number operator for mode $x$ and the second-order correlation function $G^{(2)}(a_+,b_+)$ reads
\begin{eqnarray}
&{}& G^{(2)}(a_+,b_+) =\langle\Psi |{:}\, \mathcal{N}_{a_+} \mathcal{N}_{b_+} {:}|\Psi \rangle 
\label{prob-true}
\\
\!\!&=& \!\! \sinh^2{K} \left( \sinh^2{K} + \cosh^2{K} \sin^2 (\Delta/2) \right), 
\nonumber            
\end{eqnarray}
where $\Delta = \phi_a - \phi_b$. For details of the calculation see Appendix A or~\cite{Laskowski2009}. We turn $G^{(2)}$ into the normalized second-order coherence $g^{(2)} = G^{(2)}/(\langle \mathcal{N}_{a+} \rangle \langle \mathcal{N}_{b+} \rangle)$ to compare our results with known results in literature. For $|\Psi\rangle$ we have $\langle \mathcal{N}_{a+} \rangle = \langle \mathcal{N}_{b+} \rangle = \sinh^2{K}$ and in this case
\begin{equation}
g^{(2)} = 1+ \sin^2 (\Delta/2) + \frac{1}{\sinh^2 K} \sin^2 (\Delta/2).
\end{equation}
If $\Delta=\pi$, the modes $a_+$ and $b_+$ carry orthogonal polarizations. Since such polarizations were initially correlated, see Eq.~(\ref{state-product}), e.g. $a_H$ and $b_V$, $g^{(2)}$ takes the maximal value $g^{(2)}_{max} = 2 + \frac{1}{\sinh^2 K}$~\cite{Masha2004}. If $\Delta=0$, the modes carry the same polarization and since such parallel polarizations were not correlated, e.g. modes $a_H$ and $b_H$ were independent, $g^{(2)}$ takes the minimal value $g^{(2)}_{min} = 1$. Therefore, the second-order interference visibility reads
\begin{equation}
V_2 = \frac{g^{(2)}_{max} - g^{(2)}_{min}}{g^{(2)}_{max} + g^{(2)}_{min}}
= \frac{1}{1 + 2\tanh^2 K}, 
\label{v-pdf}
\end{equation}
We show the dependence of the visibility on the amplification gain $K$ in Fig.~\ref{fig:v}. 
For small values of $K \to 0$, visibility approaches unity, the value for a two-photon singlet state. The full analogy with a two-photon singlet state indicates that the two modes can be perfectly distinguished locally, but the system maintains a superposition between these states. If the intensity of the pumping power increases, higher order emissions appear and the visibility decreases. The critical value of gain, above which visibility is not sufficient to violate the CHSH inequality equals $K_{crit}=0.49$. This corresponds to the average photon number in emissions equal to $0.26$. For this value, probability that the source emits exactly one pair of photons reads $0.13$. For large $K$, the visibility approaches $1/3$. This is the value for a thermal state. 

\subsection{``On-off'' binary detection}

Now, we consider a photon counting detector, which is denoted in Fig.~\ref{setup} by $T$. Usually, these are avalanche photodiodes, but photomultiplier may work in this regime as well. Such detector gets saturated already by a single photon input and has high quantum efficiency. Thus, effectively it discriminates only between vacuum $|0\rangle$ and non-vacuum input by a click. In principle, it is described by a POVM of the form 
\begin{equation}
\mathbf{1} - |0\rangle\langle 0|.
\end{equation}
The unit operator is $\mathbf{1} = \sum_{n=0}^{\infty} |n\rangle\langle n|$, where $|n\rangle$ denotes the photon number Fock state with $n$ photons in the observed mode.

In this model, probability that the detector will click is $1$, if there is at least one photon in the mode observed by the detector. Probability of joint detection by two detectors $T_{a_+}$ and $T_{b_+}$ in Fig.~\ref{setup} equals
\begin{eqnarray}
\label{prob-diode}
p(T_{a_+}, T_{b_+}| \phi_a,\phi_b) \!\!&=&\!\! \sum_{p,k=1}^\infty \sum_{j,l=0}^\infty |\langle p,j,k,l| \Psi \rangle_{\pm}|^2,
\end{eqnarray} 
where $|p,j,k,l\rangle$ denotes $p$, $j$, $k$, $l$ photons in modes emerging from polarization analyzers $a_+$, $a_-$, $b_+$ and $b_-$ respectively, and $|\Psi \rangle_{\pm}$ is the state $|\Psi \rangle$ expressed in this new polarization basis. One has
\begin{align}
  p=&
  1-\frac{2}{\cosh^2 K} + \frac{1}{\cosh^4 K} \frac{1}{1-\tanh^2 K\sin^2(\Delta/2)}.
  \label{eq:p}
\end{align}
The details of the calculation are given in Appendix B. Using the same method as for Eq.~(\ref{v-pdf}), we evaluated visibility by obtaining $p_{max}$ and $p_{min}$ 
\begin{equation}
V_2 = \frac{1}{2 \cosh^2 K -1}.
\end{equation}
We compared visibilities obtained with the linear efficiency and binary detections in Fig.~\ref{fig:v}. For the latter, visibility decreases in presence of higher-number contributions even faster than for linear devices. For $K > 0.44$, CHSH inequality violation is impossible. This result is not surprising at all, since the strategy based on applying detection designed for single photon counting to high intensities cannot work. However, we will show in the following Sections, that binary detection combined with multiports is a well-designed measurement approach and outperforms other strategies for multi-photon states.

\begin{figure}
\centering
\includegraphics[width=0.45\textwidth]{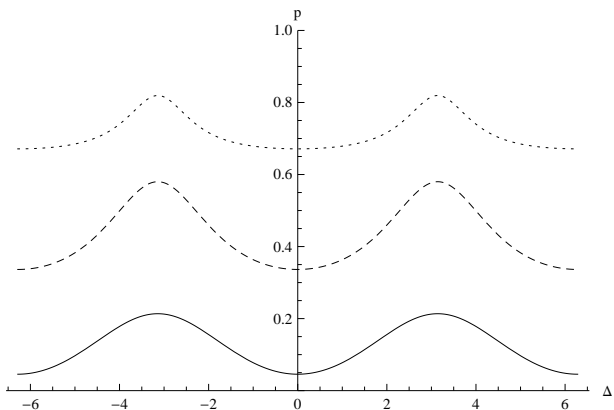}
\caption{Probability $p$ of joint detection by two detectors $T_{a_+}$ and $T_{b_+}$ in Fig.~\ref{setup} for $K=0.5$ (solid line), $K=1$ (dashed line) and $K=1.5$ (dotted line).}
\label{fig:examples}
\end{figure}

Note that the interference curve of Eq.~(\ref{eq:p}) is not a sinusoidal one. Examples of $p$ as a function of $\Delta$ are given in Fig.~\ref{fig:examples}.

\section{Hybrid detection strategy}\label{Sec2}

\begin{figure}
\centering
\includegraphics[width=6cm]{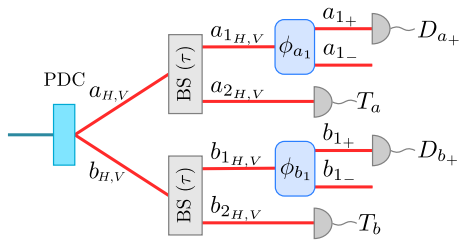}
\caption{Scheme of Fig.~\ref{setup} modified. Beam splitters coherently tap the input multi-photon state.}
\label{BS}
\end{figure}

Let us modify the setup shown in Fig.~\ref{setup} by combining the ``on-off'' detection with photon number measurements. Our new scheme is shown in Fig.~\ref{BS}. 
Two spatial modes $a$ and $b$ of state $|\Psi\rangle$ impinge on the asymmetric beam splitters (BS) with equal transmitivity coefficients $\tau \ll 1$. Since each mode contains two polarizations, mutually orthogonal, BS acts independently on these polarizations. For example, spatial mode $a_i$ is split into two modes as follows
\begin{equation}
a_i = \sqrt{\tau} a_{1,i} + \sqrt{1-\tau} a_{2,i},
\end{equation}
where $i=H,V$. One has a similar relation for mode $b$. The reflected beams $a_2$ and $b_2$ (both containing $H$ and $V$ polarizations) are sent to the ``on-off'' detectors, denoted by $T_a$ and $T_b$, respectively. These detectors conditionally project the incoming state. For this reason, our new scheme still works coherently and BSs do not introduce neither losses nor mixing of the initial state. We focus on these events where the detectors $T_a$ and $T_b$ do not register photons (the incoming intensity is zero). The transmitted beams are sent through polarization analyzers, according to transformation in Eq.~(\ref{measurement}), and their state reads
\begin{equation}
|\Psi_{\tau}\rangle = \frac{N}{\cosh^2{K}} \sum_{n=0}^{\infty} \sqrt{n+1} \, (\tau \tanh{K})^n |\psi^{(n)}_-\rangle,
\label{stateBS}
\end{equation}
where $N = \cosh^2{K} (1 - \tau^2 \tanh^2{K})$ is a new normalization constant. Now, we measure the second-order correlation function $G^{(2)}(a_{1_+},b_{1_+})$ between modes $a_{1_+}$ and $b_{1_+}$, using linear efficiency detection $D_{a_+}$ and $D_{b_+}$, in the same manner as described in Section IIA. We also calculate $G^{(2)}(a_{1_+},b_{1_+})$ as we did in Section IIA for the state given in Eq.~(\ref{state}). Since the state $|\Psi_{\tau}\rangle$ is a slight modification of the state $|\Psi\rangle$, it is easy to see that the visibility for $|\Psi_{\tau}\rangle$ equals
\begin{equation}
V_2^{(\tau)} = \frac{1}{1 + 2 \tau^2 \tanh^2{K}}.
\label{v-BS}
\end{equation}
\begin{figure}
\centering
\includegraphics[width=0.45\textwidth]{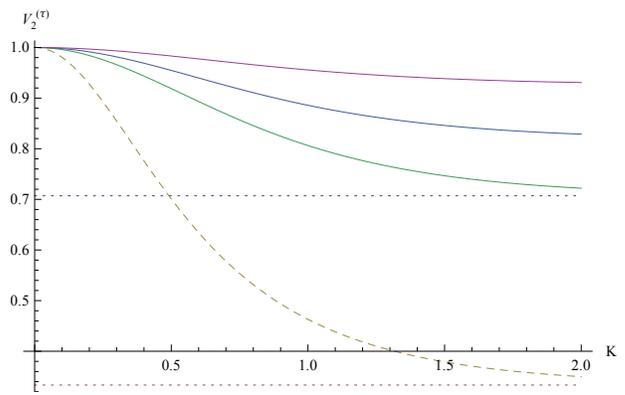}
\caption{Visibilities as functions of amplification gain $K$ for several values of transmitivity. $\tau=1$ (dashed line) corresponds to the case discussed in Section IIA. For smaller values $\tau=(1/\sqrt{2}-1/2)^{1/2},1/3,1/10$ (solid lines enumerated upwards) visibility is improved and exceeds the critical value of $1/\sqrt{2}$ (the upper horizontal dotted line), above which violation of CHSH inequality is obtained. The lower horizontal dotted line represents visibility obtained for thermal state equal $1/3$.}
\label{fig:eta-v}
\end{figure}

The visibility as a function of amplification gain $K$ is shown in Fig.~\ref{fig:eta-v} for several values of transmitivity: $\tau=1,(1/\sqrt{2}-1/2)^{1/2},1/3,1/10$. We observe large improvement of the interference contrast with respect to the cases discussed in Fig.~\ref{fig:v}. This effect is present also for very high $K$. If $\tau_{crit} = (1/\sqrt{2}-1/2)^{1/2}$, the visibility exceeds and asymptotically approaches $1/\sqrt{2}$. For all $\tau \le \tau_{crit}$, visibility exceeds this value for all $K$.

\section{Filtering by multiports} \label{Sec3}

\begin{figure}
\centering
\includegraphics[width=4.5cm]{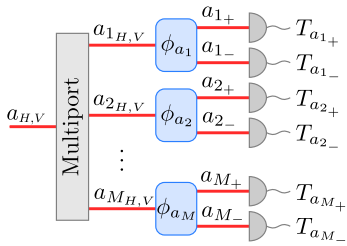}
\caption{Multiport splits incoming multi-photon beam into M spatial beams.}
\label{fig:BS2M_1}
\end{figure}

An increased interference visibility can be also obtained using multiport BS~\cite{MULTIPORT} (instead of asymmetric BS) and ``on-off'' detection only, see Fig.~\ref{fig:BS2M_1}. The action of a multiport can be thought of as ``spatial multiplexing'' and thus is similar in its spirit to time-multiplexing used in optical fibers~\cite{Fitch,Achilles}. There, the incident multi-photon pulse is split into many pulses with smaller amplitudes, displaced in time more than the dead time of the detector located at the end of the fiber. In case of a multiport, the beam is split into spatial modes, each being input to a separate detector. Ideally, each mode contains not more than a single photon.

The relation between the transmitivity $\tau$ of the asymmetric BS from Fig.~\ref{BS} and the number of ports in a multiport beam splitter $M$ is given by $\tau = 1/M$. For instance, the spatial mode $a_i$ is split into $M$ different modes as follows
\begin{equation}
a_p \to \frac{1}{\sqrt{M}}(a_{1_p} + a_{2_p} + \cdots + a_{M_p}),
\end{equation}
where $p=H,V$ and similarly for mode $b$. Next, all modes $a_{i_p}$ and $b_{i_p}$ with $i=0,..,M$, are subjected to polarization rotation given in Eq.~(\ref{measurement}). We take into account only such events where only one detector at each side, $T_{a_{i+}}$ and $T_{b_{j+}}$, clicks (there are $M^2$ such  events) and we discard all multiple counts at each side. The total probability to have single clicks at detectors observing the $+$ outputs of the polarization analyzers equals
\begin{eqnarray}
p_M = \sum_{i,j=1}^M p(T_{a_{i+}}, T_{b_{j+}}| \phi_1, \phi_2).
\end{eqnarray}
Since all probabilities in the above sum have the same value $p_M = M^2 p(T_{a_{1+}}, T_{b_{1+}} | \phi_a, \phi_b)$. The form of the state $|\Psi\rangle$ after passing the multiport and conditioned on detection of vacuum in modes $a_{2_+}, a_{2_-}...,a_{M_+}, a_{M_-}$ and $b_{2_+}, b_{2_-}...,b_{M_+}, b_{M_-}$ is given in Eq.~(\ref{stateBS}), but with $\tau$ replaced by $1/M$. Thus, we obtain $p_M = M^2 \sum_{p,k=1}^\infty \sum_{j,l=0}^\infty \left| \langle p,j,k,l| \Psi_{\frac{1}{M}}\rangle_{\pm} \right|^2$
\begin{eqnarray}
p_M \!\!\!&=&\!\! \! M^2 \!\left\{\!1 - 2\left(1-\frac{\tanh^2 K}{M^2}\right) 
+\frac{\left(1-\frac{\tanh^2 K}{M^2}\right)^2}{1-\frac{\tanh^2 K}{M^2}\sin^2(\Delta/2)} \right\}.
\nonumber \\
\end{eqnarray}
The maximal value of $p_M$ is obtained for $\Delta=\pi/2$ and minimal for $\Delta=0$ and thus, the visibility reads
\begin{equation}
V^{(M)}_2 = \frac{1-\frac{1}{M^2} \tanh^2 K}{1+\frac{1}{M^2} \tanh^2 K}.
\end{equation}

\begin{figure}
\centering
\includegraphics[width=0.45\textwidth]{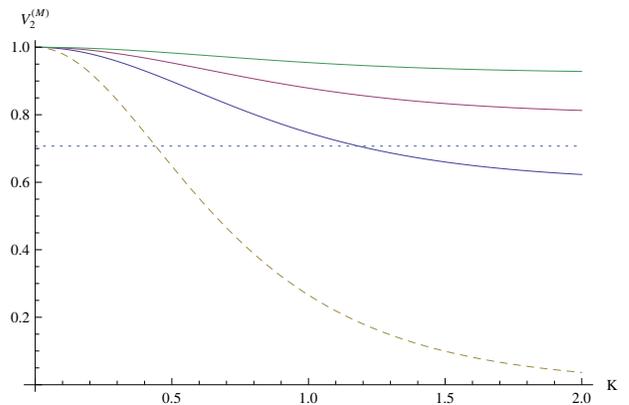}
\caption{Visibilities as functions of amplification gain $K$ for several multiports. $M=1$ (dashed line) corresponds to the case discussed in Section IIA. For multiports $M=2,3,5$ (solid lines enumerated upwards) visibility is improved and exceeds the critical value of $1/\sqrt{2}$ (horizontal dotted line), above which violation of CHSH inequality is obtained.}
\label{M-v-3}
\end{figure}

The visibility as a function of amplification gain is shown in Fig.~\ref{M-v-3} for several multiports $M=1,2,3,5$. We observe significant increase in the interference contrast with respect to the case where no conditional preparation of the state takes place. 

There is a physical explanation of this result. It also applies to the result presented in Section III. We notice that the probability $p_M$ may be expressed as follows
\begin{eqnarray}
 p_M \!\!&=&\!\!  N' \sum_{\Sigma} 
\left|\sum_{n=1}^\infty \frac{\tanh^n K }{M^{n-1}} \langle p,j,k,l \ket{\Psi^{(n)}_-}_{\pm} \right|^2,
\label{p-M}
\end{eqnarray}
where $N'$ is the normalization, $\Sigma$ denotes summation over $p,k=1,...,\infty$ and $j,l=0,...,\infty$. The summation over $n$ may be split in the following way
\begin{eqnarray}
  \tanh K \ket{\Psi^{(1)}} + \sum_{n=2}^\infty \tanh^n K \frac{1}{M^{n-1}}\ket{\Psi^{(n)}}.
\end{eqnarray}
If $M \to \infty$ only the firs term survives. It means that only creation of single pairs contributes to the probability of coincidences in case of large multiport. Effectively, it behaves like a POVM, which projects the multi-photon state on a two-photon singlet state subspace.

\section{Conclusions}

We have considered the possibility of obtaining the high interference contrast for the multi-photon states of light. Such states are produced in parametric down conversion process with strong pumping, thus they are very common and easily accessible non-classical states of photons. In the weak pumping regime, the interference visibility is high enough to obtain violation of Bell inequality. For the highly populated states, the conflict between local-realistic and quantum mechanical description is deeper, but inefficient photodetection techniques limit possibility of its observation. We showed that multiport beam splitters can be used to increase the visibility up to CHSH-Bell inequality violation in combination with both commonly used detection techniques: the analog and the photon counting devices. Effectively, they filter the two-photon singlet component out of the multi-photon state. Our scheme may be used for quantum state engineering which eliminates higher-number photon contributions. While in Ref.~\cite{MULTIPORT} it was demonstrated that multiports are useful for Bell tests involving high-dimensional systems, we have demonstrated that they can also be handy in standard interference experiments.

\acknowledgments

MS is supported by the EU 7FP Marie Curie Career Integration Grant No. 322150 ``QCAT'', NCN grant No. 2012/04/M/ST2/00789 and FNP Homing Plus project.  MZ is supported by QUASAR CHIST-REA project (NCBiR). MW is supported by QUASAR CHIST-REA and FNP Homing Plus project. WL is supported by NCN grant No. 2012/05/E/ST2/02352.

\section{Appendix A}

Here we show how Eq.~(\ref{prob-true}) was derived. First, we note that the state in Eq.~(\ref{state}) can be expressed in terms of action of two squeezing operations on vacuum
\begin{eqnarray}
|\Psi\rangle &=& \mathcal{S}|0\rangle= \mathcal{S}_1\mathcal{S}_2|0\rangle, \\
\mathcal{S}_1 &=& \exp\{-K(a_H^{\dagger} b_V^{\dagger} - h.c.) \}, \\
\mathcal{S}_2 &=& \exp\{K(a_V^{\dagger} b_H^{\dagger} - h.c.) \}.
\end{eqnarray}

In this case, it is easier to carry this computation in Heisenberg picture
\begin{equation}
\langle\Psi |{:}\, \mathcal{N}_{a_+} \mathcal{N}_{b_+} {:}|\Psi \rangle = \langle 0|{:}\, \mathcal{S}^{\dagger}\mathcal{N}_{a_+}\mathcal{S}\mathcal{S}^{\dagger} \mathcal{N}_{b_+} \mathcal{S}{:}|0 \rangle,
\end{equation}
where 
\begin{align}
\mathcal{N}_{a_+} &{}= \frac{1}{2} \left( a_H^{\dagger}a_H + a_V^{\dagger}a_V + e^{i \phi_a} a_H^{\dagger}a_V + e^{-i \phi_a} a_V^{\dagger}a_H \right), \nonumber \\
\mathcal{N}_{b_+} &{}= \frac{1}{2} \left( b_H^{\dagger}b_H + b_V^{\dagger}b_V + e^{i \phi_b} b_H^{\dagger}b_V + e^{-i \phi_b} b_V^{\dagger}b_H \right)
\nonumber
\end{align}
and using BCH formula $e^{\mathcal{A}} \mathcal{B}  e^{-\mathcal{A}} = \sum_{n=0}^{\infty}\frac{1}{n!} [A,B]^{(n)}$ we compute
\begin{align}
\mathcal{S}_1^{\dagger} a_H \mathcal{S}_1 &{}= c_K a_H + s_K b_V^{\dagger}, \,
\mathcal{S}_1^{\dagger} a_H^{\dagger} \mathcal{S}_1 = c_K a_H^{\dagger} + s_K b_V, \nonumber
\\
\mathcal{S}_2^{\dagger} a_V \mathcal{S}_2 &{}= c_K a_V - s_K b_H^{\dagger}, \,
\mathcal{S}_2^{\dagger} a_V^{\dagger} \mathcal{S}_2 = c_K a_V^{\dagger} - s_K b_H, \nonumber
\\
\mathcal{S}_1^{\dagger} b_V \mathcal{S}_1 &{}= c_K b_V + s_K a_H^{\dagger}, \,
\mathcal{S}_1^{\dagger} b_V^{\dagger} \mathcal{S}_1 = c_K b_V^{\dagger} + s_K a_H, \nonumber
\\
\mathcal{S}_2^{\dagger} b_H \mathcal{S}_2 &{}= c_K b_H - s_K a_V^{\dagger}, \,
\mathcal{S}_2^{\dagger} b_H^{\dagger} \mathcal{S}_2 = c_K b_H^{\dagger} - s_K a_V,\nonumber
\end{align}
where $c_K = \cosh K$ and $s_K = \sinh K$. The contributing terms in the product are 
\begin{align}
& \langle\Psi |{:}\, \mathcal{N}_{a_+} \mathcal{N}_{b_+} {:}|\Psi \rangle = 
\nonumber\\
& \frac{s_K^2}{4} \langle 0| \left[b_V(c_K a_H + s_K b_V^{\dagger}) - b_H(c_K a_V - s_K b_H^{\dagger})
\right. \nonumber\\
&\quad+ \left. e^{i \phi_a} b_V(c_K a_V - s_K b_H^{\dagger}) - e^{-i \phi_a} b_H(c_K a_H + s_K b_V^{\dagger})\right]
\nonumber \\
& \left[2 s_K - c_K a_V^{\dagger}b_H^{\dagger} + c_K a_H^{\dagger}b_V^{\dagger} 
\right. \nonumber \\
&\quad+ \left. e^{i \phi_b} c_K a_H^{\dagger}b_H^{\dagger} - e^{-i \phi_b} c_K a_V^{\dagger}b_V^{\dagger}\right]|0\rangle,
\nonumber
\end{align}
which after evaluating the products gives directly Eq.~(\ref{prob-true}).

\section{Appendix B}

The state $|\Psi \rangle$ expressed in new polarization basis given by the modes emerging from polarization analyzers $a_+$, $a_-$, $b_+$ and $b_-$ takes the form
\begin{align}
|\Psi\rangle_{\pm} ={}& \frac{1}{\cosh^2{K}} \sum_{n=0}^{\infty} \sqrt{n+1} \tanh^n{K} |\psi^{(n)}_-\rangle_{\pm},
\nonumber\\
|\psi^{(n)}_-\rangle_{\pm} ={}& \frac{1}{\sqrt{n+1}} \frac{(-1)^n}{2^n} \sum_{m=0}^n \frac{(-1)^m}{m! (n-m)!} e^{i m \phi_a} e^{i (n-m) \phi_b}
\nonumber\\
& \sum_{j_1,j_4 = 0}^{n-m} \sum_{j_2,j_3 = 0}^{m} C_{j_1}^{n-m}C_{j_4}^{n-m} C_{j_2}^{m} C_{j_3}^{m} (-1)^{j_2 + j_4}
\nonumber\\
& \sqrt{J_{12}! (n-J_{12})!J_{34}! (n-J_{34})!} 
\nonumber\\
& |J_{12}\rangle_{a}^+ |n-J_{12}\rangle_{a}^- |J_{34}\rangle_{b}^+ |n-J_{34}\rangle_{b}^-,
\label{statepm}
\end{align}
where $C_j^n = {n \choose j}$, $J_{xy} = j_x+j_y$, $x,y \in \{1,2,3,4\}$. 

To compute Eq.~(\ref{prob-diode}), we note that 
\begin{equation}
p = 1-p_0-p_1-p_2,
\end{equation}
where 
\begin{align}
p_0 =&\sum_{j,l=0}^\infty | \langle 0, j, 0, l | \Psi \rangle_{\pm}|^2, 
\\
p_1 =& \sum_{j,l=0}^\infty \sum_{p=1}^{\infty} |\langle p,j,0,l | \Psi \rangle_{\pm}|^2, 
\\
p_2 =& \sum_{j,l=0}^\infty \sum_{p=1}^{\infty}|\langle 0, j, p, l | \Psi\rangle_{\pm} |^2, 
\end{align}
where $p_2= p_1$, due to symmetry of the setup. 

\subsection{Evaluation of $p_0$}

\begin{align}
  & \langle 0,j,0,l\vert\psi_{-}^{(n)}\rangle_{\pm}
  =
  \frac{1}{\sqrt{n+1}}\,\frac{(-1)^n}{2^n}\,
  \sum_{m=0}^n\frac{(-1)^m}{m!\,(n-m)!}
  \nonumber\\
  & \quad
  \sum_{j_1,j_4=0}^{n-m}\sum_{j_2,j_3=0}^m
  C_{j_1}^{n-m}\,C_{j_4}^{n-m}\,C_{j_2}^m\,C_{j_3}^m\,(-1)^{j_2+j_4}
  \nonumber\\
  &\quad
  \sqrt{J_{12}!\,(n-J_{12})!\,J_{34}!\,(n-J_{34})!}\, e^{im\varphi_a}\,e^{i(n-m)\varphi_b}
  \nonumber\\
  &\quad
  \delta_{0=J_{12}}\,
  \delta_{j=n-J_{12}}\,
  \delta_{0=J_{34}}\,
  \delta_{l=n-J_{34}}.
\end{align}
We note that the Kronecker deltas impose the following conditions $j_1=0, j_2=0, j_3=0, j_4=0, j=n, l=n$ and $C_0^{n}=1$, which simplify the above formula to
\begin{align}
  & \langle 0,j,0,l\vert\psi_{-}^{(n)}\rangle_{\pm}
  =
  \frac{1}{\sqrt{n+1}}\,\frac{(-1)^n}{2^n}\,
  \sum_{m=0}^n\frac{(-1)^m}{m!\,(n-m)!}
  \nonumber\\
  &\qquad e^{im\varphi_a}\,e^{i(n-m)\varphi_b}\,
  n!\,
  \delta_{j=n}\,
  \delta_{l=n}
  \nonumber\\
  &\quad = \frac{1}{\sqrt{n+1}}\, (i e^{i (\phi_a + \phi_b)/2} \sin(\Delta/2))^n \,
  \delta_{j=n}\, \delta_{l=n}, 
\end{align}
\begin{align}  
  &\langle 0,j,0,l\vert\Psi\rangle_{\pm} =
  \frac{1}{\cosh^2 K}
  \left(\tanh K\,i e^{i(\varphi_a+\varphi_b)/2}\sin(\Delta/2)\right)^j
  \delta_{l=j},
\end{align}  
\begin{align}  
  &|\langle 0,j,0,l\vert\Psi\rangle_{\pm}| =
  \frac{1}{\cosh^2 K}
  \left(\tanh K\,\sin(\Delta/2)\right)^j \delta_{l=j}.
\end{align}  
Finally,
\begin{align}
  & p_0
   =
  \frac{1}{\cosh^4 K} \sum_{j=0}^{\infty}
  \left\lvert\tanh K\,\sin(\Delta/2)\right\rvert^{2j}
  \nonumber\\
  &
  = \frac{1}{\cosh^4 K}\, \frac{1}{1-\tanh^2 K\,\sin^2(\Delta/2)}.
\end{align}

\subsection{Evaluation of $p_1$}

\begin{align}
  & \langle p,j,0,l\vert\psi_{-}^{(n)}\rangle_{\pm}
  =
  \frac{1}{\sqrt{n+1}}\,\frac{(-1)^n}{2^n}\,
  \sum_{m=0}^n\frac{(-1)^m}{m!\,(n-m)!}
  \nonumber\\
   &\quad
  \sum_{j_1,j_4=0}^{n-m}\sum_{j_2,j_3=0}^m
  C_{j_1}^{n-m}\,C_{j_4}^{n-m}\,C_{j_2}^m\,C_{j_3}^m\,(-1)^{j_2+j_4}
  \nonumber\\
   &\quad
  \sqrt{J_{12}!\,(n-J_{12})!\,J_{34}!\,(n-J_{34})!}\, e^{i m\varphi_a}\,e^{i(n-m)\varphi_b}
  \nonumber\\
   &\quad
  \delta_{p=J_{12}}\,
  \delta_{j=n-J_{12}}\,
  \delta_{0=J_{34}}\,
  \delta_{l=n-J_{34}}.
\end{align}
The Kronecker deltas impose the following conditions $j_1+j_2=p, j_3=0, j_4=0, j=n-p, l=n$ which simplify the formula to
\begin{align}
  & \langle p,j,0,l\vert\psi_{-}^{(n)}\rangle_{\pm}
   =
  \frac{1}{\sqrt{n+1}}\,\frac{(-1)^n}{2^n}\,
  \sqrt{p!\,(n-p)!\,n!}
  \nonumber\\
  &\quad
  \delta_{j=n-p}\,\delta_{n=l}\, 
  \sum_{m=0}^n\frac{(-1)^m}{m!\,(n-m)!}\,e^{i m\varphi_a}\,e^{i(n-m)\varphi_b}
  \nonumber\\
  &\quad
  \sum_{j_1=0}^{n-m}\sum_{j_2=0}^m
  C_{j_1}^{n-m}\,C_{j_2}^m\,(-1)^{j_2}
  \delta_{p=j_1+j_2},
\end{align}

\begin{align}
  & \langle p,j,0,l\vert\Psi\rangle_{\pm}
  =
  \frac{1}{\cosh^2 K}
  \sum_{n=0}^{\infty}\,\tanh^n K\,
  \frac{(-1)^n}{2^n}\,
  \sqrt{p!\,(n-p)!\,n!}
  \nonumber\\
  &\qquad
  \delta_{j=n-p}\,
  \delta_{n=l}
  \sum_{m=0}^n\frac{(-1)^m}{m!\,(n-m)!}\,e^{im\varphi_a}\,e^{i(n-m)\varphi_b}
  \nonumber\\
  &\qquad
  \sum_{j_1=0}^{n-m}\sum_{j_2=0}^m
  C_{j_1}^{n-m}\,C_{j_2}^m\,(-1)^{j_2}
  \delta_{p=j_1+j_2}
  \\
  &\quad=
  \label{eq:psi2}
  \frac{1}{\cosh^2 K}
  \tanh^l K\,
  \frac{1}{2^l}\,(-1)^l\,
  \sqrt{p!\,j!\,l!}\,
  \delta_{j=l-p}\,
  \nonumber\\
  &\qquad
  \sum_{m=0}^l
  \sum_{j_1=0}^{l-m}\sum_{j_2=0}^m
  \frac{e^{i m\varphi_a}\,e^{i(l-m)\varphi_b}\,(-1)^{m+j_2}}{j_1!\,(l-m-j_1)!\,j_2!\,(m-j_2)!}
  \delta_{p=j_1+j_2}.
\end{align}
We change of the order of sums
\begin{align}
  &\sum_{m=0}^l\sum_{j_1=0}^{l-m}\sum_{j_2=0}^m f(m,j_1,j_2)\,\delta_{p=j_1+j_2}
  =
  \nonumber\\
  &\quad =
  \sum_{j_1=0}^l\sum_{j_2=0}^l\sum_{m=j_2}^{l-j_1}f(m,j_1,j_2)\,\delta_{p=j_1+j_2}
  \nonumber\\
  &\quad =
  \sum_{j_1=0}^p\sum_{m=p-j_1}^{l-j_1}f(m,j_1,p-j_1)
  \nonumber\\
  &\quad =
  \sum_{j_1=0}^p\sum_{m=p}^{l}f(m-j_1,j_1,p-j_1).
\end{align}
One has to remember also that $p\le l$.  After the above change
\begin{align}
  & \langle p,j,0,l\vert\Psi\rangle_{\pm}
  =
  \frac{1}{\cosh^2 K}
  \tanh^l K\,
  \frac{1}{2^l}\,(-1)^l\,
  \sqrt{p!\,j!\,l!}\,
  \delta_{j=l-p}\,
  \nonumber\\&\qquad
  \left(
  \sum_{j_1=0}^p
  \frac{e^{ij_1(\varphi_b-\varphi_a)}}{j_1!\,(p-j_1)!}
  \right)
  \left(
  e^{i p\varphi_a}
  \sum_{m=0}^{l-p}
  \frac{e^{i m\varphi_a}\,e^{i(l-p-m)\varphi_b}\,(-1)^m}{(l-p-m)!\,m!}
  \right),
  \nonumber\\&\quad
   =
  \frac{1}{\cosh^2 K}
  \tanh^l K\,
  (-1)^l\,
  \sqrt{l!}\, e^{i l(\varphi_a+\varphi_b)/2}
  \nonumber\\&\qquad
  \frac{\left(\cos(\Delta/2)\right)^p}{\sqrt{p!}}
  \frac{\left(i \sin(\Delta/2)\right)^{l-p}}{\sqrt{(l-p)!}}
  \,
  \delta_{j=l-p},
\end{align}
\begin{align}
  & |\langle p,j,0,l\vert\Psi\rangle_{\pm}|
   =
  \frac{1}{\cosh^2 K}
  \tanh^l K\,
  \sqrt{l!}
  \nonumber\\&\quad
  \frac{\left(\cos(\Delta/2)\right)^p}{\sqrt{p!}}
  \frac{\left( \sin(\Delta/2)\right)^{l-p}}{\sqrt{(l-p)!}}
  \,
  \delta_{j=l-p},
\end{align}
where we used
\begin{align}
  \sum_{j_1=0}^p
  \frac{e^{i j_1\Delta}}{j_1!\,(p-j_1)!}
  &=
  \frac{\left(e^{i\Delta}+1\right)^p}{p!},
  \\
  \sum_{m=0}^{l-p}
  \frac{e^{i m\varphi_a}\,e^{i(l-p-m)\varphi_b}\,(-1)^m}{(l-p-m)!\,m!}
  &=
  \frac{\left(e^{i\varphi_b}-e^{i\varphi_a}\right)^{l-p}}{(l-p)!}.
\end{align}
Finally, 
\begin{align}
  p_1
  ={}&
  \frac{1}{\cosh^4 K}
  \sum_{l=0}^{\infty}\tanh^{2l} K\sum_{p=1}^{l}
  \binom{l}{p}
  \cos^{2i}(\tfrac{\Delta}{2})
  \sin^{2(l-p)}(\tfrac{\Delta}{2})
  \nonumber\\
  ={}&
  \frac{1}{\cosh^4 K}
  \sum_{l=0}^{\infty}\tanh^{2l} K\left(1-\sin^{2l}(\Delta/2)\right)
  \nonumber\\
  ={}&
  \frac{1}{\cosh^4 K}\left( \cosh^2 K - \frac{1}{1-\tanh^2 K\sin^2(\Delta/2)}\right).
\end{align}

\end{document}